\documentclass{natureprintstyle} 

\usepackage{xcolor}
\usepackage{graphicx}
\usepackage{dcolumn}
\usepackage{bm}
\usepackage{color}
\usepackage{hyperref}
\usepackage{units}
\usepackage{hhline}
\usepackage{amsmath,amssymb}
\usepackage{float}

\newcommand{\B}{\mathcal{B}}

\newcommand{\ninetyrisetime}{\unit[12.6]{s}}
\newcommand{\ninetyrisetimelogistic}{\unit[7.64]{s}}
\newcommand{\fzeroref}{\unit[11.186433]{Hz}}
\newcommand{\mjdref}{57734.4849906}
\newcommand{\slidingwindowlength}{\unit[200]{s}}

\newcommand{\stepglitchdf}{\unit[{16.11}_{-0.04}^{+0.04}]{\mu Hz}}
\newcommand{\stepglitchtg}{\unit[{-0.31}_{-2.78}^{+2.74}]{s}}

\newcommand{\htwotaud}{\unit[{65.97}_{-24.44}^{+59.38}]{s}}
\newcommand{\htwodeltafdec}{\unit[{9.36}_{-6.38}^{+12.20}]{\mu Hz}}

\newcommand{\hptaudec}{\unit[{53.96}_{-14.82}^{+24.02}]{s}}
\newcommand{\hpdeltafdec}{\unit[{17.77}_{-7.99}^{+13.68}]{\mu Hz}}

\newcommand{\hpdeltafpre}{\unit[{5.40}_{-2.05}^{+3.39}]{\mu Hz}}
\newcommand{\hpdeltaf}{\unit[{16.01}_{-0.05}^{+0.05}]{\mu Hz}}

\newcommand{\SPA}{School of Physics and Astronomy, Monash University, VIC 3800, Australia}
\newcommand{\OzGravMonash}{OzGrav: The ARC Centre of Excellence for Gravitational Wave Discovery, Clayton VIC 3800, Australia}

\usepackage{siunitx} 
\ExplSyntaxOn
\NewDocumentCommand{\setvariable}{mmm}
 {
  \prop_gclear_new:c { g_giacomo_var_#1_prop }
  \prop_gput:cnn { g_giacomo_var_#1_prop } { value } { #2 }
  \prop_gput:cnn { g_giacomo_var_#1_prop } { unit } { #3 }
 }

\NewDocumentCommand{\printvariable}{sm}
 {
  \IfBooleanTF{#1}
   {
    \num { \__giacomo_get:nn { #2 } { value } }
   }
   {
    \exp_args:Nnx \SI { \__giacomo_get:nn { #2 } { value } }
        { \__giacomo_get:nn { #2 } { unit } }
   }
 }

\DeclareExpandableDocumentCommand{\getvalue}{m}
 {
  \__giacomo_get:nn { #1 } { value }
 }
\cs_new:Npn \__giacomo_get:nn #1 #2
 {
  \prop_item:cn { g_giacomo_var_#1_prop } { #2 }
 }
\cs_set_eq:NN \fpeval \fp_eval:n
\ExplSyntaxOff

\newcommand{\bayesf}[2]{\num[round-mode=figures,round-precision=2]{\fpeval{((\getvalue{#1} - \getvalue{#2}) / 2.302585)}}}

\setvariable{step}{319099.970}{}
\setvariable{one_exponential}{319096.005}{}
\setvariable{two_exponential}{319100.836}{}
\setvariable{antiglitch}{319106.532}{}
\setvariable{logistic}{319095.05}{}
\setvariable{step_decay}{319104.5752137798}{}

\title{Rotational evolution of the Vela pulsar during the 2016 glitch}

\author{Gregory Ashton$^{1, 2}$, Paul D. Lasky$^{1, 2}$, Vanessa Graber$^{3}$, Jim Palfreyman$^{4}$}

\begin{document}
\maketitle

\begin{affiliations}
\item \SPA
\item \OzGravMonash
\item Department of Physics and McGill Space Institute, McGill University, 3550 rue University, Montreal QC H3A 2T8, Canada
\item School of Natural Sciences,  University of Tasmania, Australia
\end{affiliations}

\begin{abstract}
The 2016 Vela glitch observed by the Mt Pleasant radio telescope provides the first opportunity to study pulse-to-pulse dynamics of a pulsar glitch, opening up new possibilities to study the neutron star's interior. 
We fit models of the star's rotation frequency to the pulsar data, and present three new results.  First, we constrain the glitch rise time to less than \ninetyrisetime{} with 90\% confidence, almost three times shorter than the previous best constraint. Second, we find definitive evidence for a rotational-frequency overshoot and fast relaxation following the glitch. Third, we find evidence for a slow-down of the star's rotation immediately prior to the glitch.  The overshoot is predicted theoretically by some models; we discuss implications of the glitch rise and overshoot decay times on internal neutron-star physics. The slow down preceding the glitch is unexpected; we propose the slow-down may trigger the glitch by causing a critical lag between crustal superfluid and the crust.
\end{abstract}

Pulsar glitches, rotational irregularities of otherwise stably rotating neutron stars, are believed to be caused by the complex interplay between micro- and macrophysical properties of the star's internal components. One model posits that superfluid vortices in the inner crust suddenly unpin, transferring angular momentum to the star's lattice crust\cite{anderson75}.  This is seen as an increase in the frequency of pulsations.  Such models are difficult to verify; the internal components of the star are shielded from view and their behavior has to be inferred indirectly. Until recently, radio observations of glitches were limited to observations before and after the glitch, but not during. The detailed morphology of glitch dynamics (e.g., the glitch rise time) is therefore not well constrained or understood. In 2016, the first pulse-to-pulse observations of a glitch were made using the University of Tasmania Mt Pleasant \unit[26]{m} radio telescope\cite{palfreyman18}. Those observations showed variations in pulse shape of four pulses starting 20 rotations before the inferred time of the glitch, which are attributed to variations in the magnetospheric state. A preliminary estimate of $\sim \unit[4.4]{s}$ for the glitch rise time was given.

In this Article, we provide a detailed pulse-to-pulse analysis of the glitch morphology.  Our main results are three-fold.
First, we constrain the glitch rise time to less than \ninetyrisetime{} with 90\% confidence.  We connect this with internal neutron-star physics using a body-averaged description of the components participating in the glitch. Second, we show that a frequency overshoot ---an increase in the rotation frequency above the post-glitch equilibrium value --- and subsequent fast relaxation exist immediately following the glitch, in agreement with the only two previous high time-resolution observations of Vela glitches\cite{dodson2002, dodson07}, and as explained by several models of neutron-star glitches\cite{vaneysden2010,haskell12,antonelli17,graber18}.  
Third, we show that the glitch may be preceded by an initial precursor slow-down, whereby the crust of the star slowed before rapidly speeding up; we speculate that this preceding slow-down of the pulsar's rotation triggered the glitch.

\begin{figure}
    \centering
    \includegraphics[width=\columnwidth]{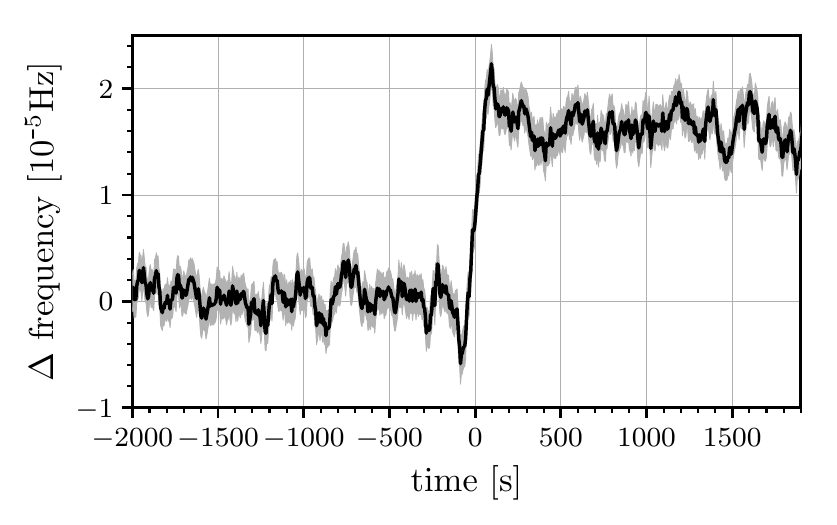}
  \caption{Rotational frequency evolution: we fit a constant-frequency model in \slidingwindowlength-long sliding windows. The frequency and time are given relative to nominal values;
  see text for details. The sliding window elongates features with respect to their true temporal evolution, and suppresses stochastic fluctuations of the frequency.  Despite this, the largest fluctuations can be seen immediately following the glitch (the overshoot) and immediately prior to the glitch (the precursor slow-down). \label{fig:frequency_evolution_only}}
\end{figure}

In the superfluid vortex model of pulsar glitches, the star's crust, and components tightly coupled to it, spin down due to external electromagnetic torques.  The crustal superfluid is decoupled from the lattice as a result of vortex pinning, implying a lag develops between the superfluid's angular velocity $\Omega_{\rm sf}$ and the crust's angular velocity $\Omega_{\rm crust}$; $\Omega_{\rm lag}\equiv\Omega_{\rm sf}-\Omega_{\rm crust}>0$.  When $\Omega_{\rm lag}$ reaches a critical value $\Omega_{\rm crit}$, some mechanism initiates the glitch, simultaneously releasing a large number of vortices. The superfluid's excess angular momentum is transferred to the crust, producing the observed spin up. The specific trigger for the angular-momentum transfer is not well understood, although speculation abounds\cite{ruderman1976, cheng88, alpar1994, andersson03, glampedakis09, peralta09, warszawski13, andersson13}.

We show there exist fluctuations of Vela's spin period prior to the glitch, see Fig.~\ref{fig:frequency_evolution_only}, and speculate that $\Omega_{\rm lag}>\Omega_{\rm crit}$ is reached due to a single, relatively large stochastic fluctuation.  This is independent of the glitch trigger mechanism, but provides a means for reaching the critical lag between superfluid and rigid crust; we discuss implications below.


\section{Frequency evolution}
For a model-agnostic view of the pulsar's evolution, we fit\cite{hobbs2006,edwards2006,bilby} a constant-frequency model to \slidingwindowlength-long data segments, sliding this window throughout the data (for details, see the Methods section). In Fig.~\ref{fig:frequency_evolution_only} we show the median frequency and 90\% credible interval for each window.  The frequency evolution on the vertical axis is given as the difference between the inferred frequency and a nominal value of \fzeroref.  Times are relative to the solar-system barycentre time of the fitted glitch MJD \mjdref{}\cite{palfreyman18}. As we detail below, this analysis method allows for a powerful comparison with physical models of pulsars' rotational evolution. 
The glitch can clearly be seen near time zero. We caution that one cannot use this plot to find the glitch rise time as the sliding window time averages the frequency evolution, thereby elongating such features.  
The time window also suppresses the amplitude of features whose timescale is shorter than the window length.  Despite this, Fig.~\ref{fig:frequency_evolution_only} shows stochastic fluctuations in the frequency evolution for the $\sim\unit[2000]{s}$ before and after the glitch.  The largest fluctuations are immediately before, and immediately after the glitch.  This motivates the more thorough analysis detailed below. 

To quantitatively analyse the frequency evolution, we define three models. The most general of these is $H_2$, where rotational changes are characterised by a constant term plus \emph{two} exponentials,
\begin{align}
    f_2 (t) = f_0 + H(t {-} t_g) \left[ \Delta f +  \Delta f_r e^{\frac{-(t-t_g)}{\tau_r}}
    + \Delta f_d e^{\frac{-(t-t_g)}{\tau_d}}\right],
    \label{eqn-H2}
\end{align}
where $H(t - t_g)$ is the Heaviside step function, $t_g$ is the glitch time, and $\Delta f, \Delta f_r, \Delta f_d$ are the glitch magnitude and amplitudes of each exponential term. Equation~\eqref{eqn-H2} is supplemented by the relation $\Delta f + \Delta f_r + \Delta f_d = 0$ ensuring the frequency evolution is continuous at the glitch. This relation, along with positive log-uniform priors on $\Delta f$ and $\Delta f_d$ imply $\Delta f_r \le 0$; in Eq.~\eqref{eqn-H2}, the first exponential therefore describes the ``rise'' in the frequency on time-scale $\tau_r$, while the second exponential describes a ``decay'' in frequency on time-scale $\tau_d$.

Models $H_s$ and $H_1$, specified in Tab.~\ref{tab:models}, are limiting cases of $H_2$. These phenomenological models are motivated by analytic solutions to coupled rigid-body problems, discussed later where we also introduce a final model $H_{2+p}$.

For each model, we integrate the frequency evolution to obtain the phase evolution, which we invert to obtain the model-predicted arrival time for each pulse.  A likelihood of the model given the data is calculated by modelling the pulse arrival time as a sum of the deterministic arrival time predicted by the timing model, and a zero-mean Gaussian process with unknown variance.  Using this likelihood and a suitable set of priors for the model parameters, we infer their posterior distributions and the evidence for the model using {\tt PyMultiNest}\cite{buchner2014, multinest1,multinest2}. Complete descriptions of the likelihood and prior are given in the Methods.

\begin{table}
\center
\begin{tabular}{l|l|r}
 model & $f(t)$ parameters & $\log_{10} B$\\ \hline
$H_s$ & $\Delta f_r = \Delta f_d = 0$ & --- \\
$H_1$ & $\Delta f_r \neq 0 $, $\Delta f_d = 0$  & \bayesf{one_exponential}{step} \\
$H_2$ & $\{ \Delta f_r, \Delta f_d \} \neq 0 $ & \bayesf{two_exponential}{step} \\
$H_{2+p}$ & see Eq.~\eqref{eqn-Hp} & \bayesf{antiglitch}{step}
\end{tabular}
\caption{Definitions and Bayes factors for the four primary models tested in this
Article, see Eq.~\eqref{eqn-H2} and the consistency relation. The
final column is the log-Bayes factor between each model and the
step-glitch model $H_s$. Uncertainties on the log-Bayes factors are~$\lesssim 0.2$.}
\label{tab:models}
\end{table}


The simplest model in Tab.~\ref{tab:models}, $H_s$---\emph{step glitch}, ignores the complex morphology of the glitch and models the frequency evolution as a simple step function of amplitude $\Delta f$ at time $t_g$. We use this as a base-model against which we compare all other models in Tab.~\ref{tab:models}. For the $H_s$ model, we infer a glitch magnitude $\Delta f=\stepglitchdf$ and time, $t_g=\stepglitchtg$ (given relative to the reported value) consistent with the initial obsevation\cite{palfreyman18}.

\section{Glitch rise time}
We analyse the rise time using a simple, physically-motivated reference model;
a body-averaged model with two uncoupled spinning components that suddenly couple, has equations of motion that can be integrated to give a model with a single exponential rise time $\tau_r$\cite{sidery2010}, corresponding to model $H_1$ in Tab.~\ref{tab:models}. In the limit where $\tau_r$ is much smaller than the pulse period, the $H_1$ model is equivalent to the step-glitch model $H_s$.

\begin{figure}
\centering
    \includegraphics[width=\columnwidth]{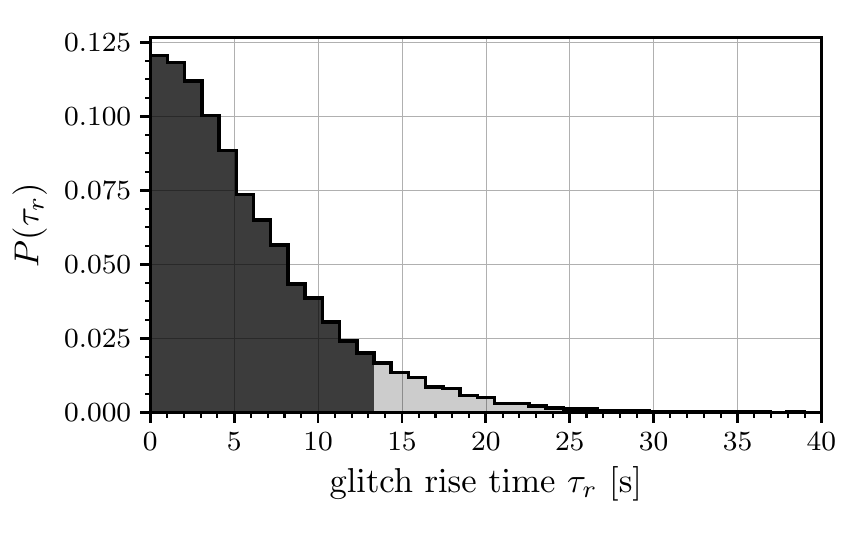}
  \caption{Posterior distribution $p(\tau_r)$ for the glitch rise time, $\tau_r$.  The dark region marks the 90\% confidence interval $\tau_r\le \ninetyrisetime$.  The distribution peaks at zero, consistent with the Bayes factor which supports the simpler $H_s$ model.  \label{fig:tau_r_posterior}}
\end{figure}

In Fig.~\ref{fig:tau_r_posterior}, we show the $\tau_r$ posterior for the $H_1$  model. The posterior peaks at zero; we cannot resolve the rise time of the glitch. This is consistent with the Bayes factor between $H_1$ and $H_s$ being in favour of the simpler step-glitch model; Tab.~\ref{tab:models}. Nevertheless, the $\tau_r$ posterior gives a 90\% upper limit of $\tau_r\le \ninetyrisetime$. This improves upon the previous best upper limit on the rise time of the 2004 Vela glitch of $\tau_r\lesssim\unit[30]{s}$\cite{dodson07}.

We also fit a model with a logistic function such that $f(t)\propto1/\left(1+e^{-t/\tau_r}\right)$. While the functional form differs, it captures a similar idea of a rise in frequency. Results are similar to the $H_1$ model; the Bayes factor favours the step glitch, with an upper limit $\tau_r\le$ \ninetyrisetimelogistic.  The evidence for the logistic and $H_1$ models are comparable, implying neither is substantially favoured by the data. The physically-motivated reference model $H_1$ and the logistic model are phenomenological; we expect the true evolution of the glitch rise to be more complicated, although we know of no robust predictions in the literature.  Nevertheless, the use of these two models shows our derived rise time is relatively insensitive to the details of the mathematical model.  Throughout this work, we quote the more conservative upper-limit rise time of the $H_1$ model.

Within body-averaged models, the glitch rise is described by a dimensionless mutual-friction coefficient $\B$, controlled by the underlying vortex dynamics\cite{alpar84, mendell91, andersson06, graber18}. Invoking a simplified two-component model, where the superfluid in the inner crust provides the angular-momentum reservoir for the glitch, the rise time is an indirect measurement of the coupling strength between the superfluid and the crust, with moments of inertia $I_{\rm sf}$ and $I_{\rm crust}$, respectively. We derive a lower limit
\begin{align}
    \B 
    \gtrsim 5.7\times 10^{-6} \left(\frac{\tau_r}{\ninetyrisetime}\right)^{-1} \left(\frac{f_{\rm sf}}{\unit[11]{Hz}}\right)^{-1} \left(\frac{ I_{\rm sf}/I_{\rm tot}}{0.01}\right),
\label{eqn:B}
\end{align}
where $f_{\rm sf}$ is the rotation frequency of the superfluid and $I_{\rm tot}= I_{\rm sf} + I_{\rm crust}$.


\section{Glitch overshoot and relaxation}
To investigate the overshoot and subsequent relaxation, we include a second exponential in the frequency evolution. This $H_2$ model (Tab.~\ref{tab:models}) is a simplified version of the three-component neutron-star model from Ref.\cite{graber18}, where the star is separated into crustal superfluid, core superfluid and the non-superfluid crust component.  The $H_2$ model assumes the three constituents are rigidly rotating and coupled via constant mutual-friction coefficients. While the specific equation chosen to model the overshoot and decay is motivated by this three-component model, we treat it as phenomenological for understanding the glitch dynamics. This phenomenological model could also be interpreted in terms of alternative physical models that also predict frequency overshoots\cite{vaneysden2010,haskell12,antonelli17}.

\begin{figure}
\centering
    \includegraphics[width=\columnwidth]{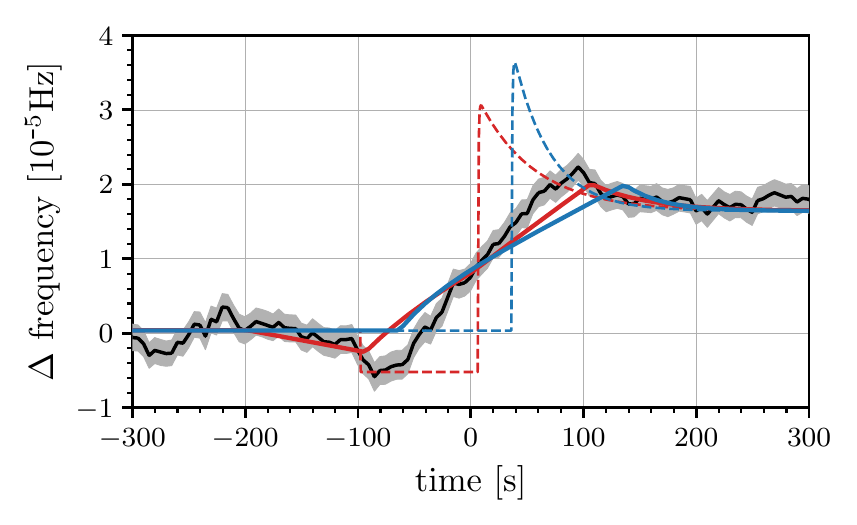}
  \caption{Rotational frequency evolution of the data (black and grey; reproduced from Fig.~\ref{fig:frequency_evolution_only}) and best-fit models.  We show the maximum-likelihood fit for the model that includes a step glitch with an overshoot and subsequent exponential decay ($H_{2}$; blue curves), and for a model that includes a slow-down preceding the glitch ($H_{2+p}$; red curves).  Dashed curves show the raw frequency evolution. Solid curves shows the time-averaged frequency evolution, which can be compared directly with the time-averaged data (black).   \label{fig:frequency_evolution}}
\end{figure}

We fit model $H_2$ to the data, and show the maximum likelihood alongside the time-windowed data in Fig.~\ref{fig:frequency_evolution}. We show the raw model (dashed blue curve) and the time-averaged frequency evolution (solid blue curve); the latter can be directly compared to the time-windowed data (black curve). 

Comparing the overshoot-decay model $H_{2}$ and step-glitch model $H_s$ yields a Bayes factor $\log_{10} B=\bayesf{two_exponential}{step}$, providing marginal support for the overshoot model.  However, the $H_1$ model (which compared unfavourably against $H_s$) is a special case of model $H_2$.  The more relevant Bayes factor to understand the importance of the overshoot and relaxation is between $H_2$ and $H_1$, for which $\log_{10}B= \bayesf{two_exponential}{one_exponential}$ showing substantial evidence in favour of the overshoot and relaxation. Alternatively, we can compare $H_s$ with a modified step-function evolution including a single decaying exponential. This model, often used in glitch-timing to model long-term, $\mathcal{O}(\gtrsim \unit[1]{day})$, relaxation was also fit to the data: the Bayes factor, $\log_{10} B = \bayesf{step_decay}{step}$, demonstrates strong support for a relaxation component with $\tau_d \sim \unit[1]{min}$, further confirming the existence of the overshoot. We remind the reader that a Bayes factor of $\log_{10}B>2$ is considered ``decisive'' support for a model, while $1\le\log_{10}B\le2$ is considered ``strong'' support\cite{kass1995}.

The existence of the overshoot is clear both visually and through our quantitative analysis.  This is \emph{not} the first identification of an overshoot: Refs.\cite{dodson2002,dodson07} found a similar feature in the 2000 and 2004 Vela glitches, the only other pulsar glitches with high-time resolution data. However, these were not as well resolved as the telescope was less sensitive, requiring \unit[10]{s}-folding of the pulses to achieve sufficient signal-to-noise ratio to calculate times of arrival. Comparing with the work herein, the pulse folding also likely explains the less constrained $\lesssim$\unit[30]{s} glitch rise time.

The maximum-likelihood overshoot-decay model $H_{2}$ shown in Fig.~\ref{fig:frequency_evolution} has a decay timescale of $\tau_d{=}\htwotaud$, magnitude $\Delta f_d{=}\htwodeltafdec$, and the rise time is similarly constrained as in the $H_1$ model. The large uncertainty on the decay time is due to a strong correlation with the size of the overshoot: larger overshoots with shorter decay times are just as probable as smaller overshoots with longer decay times.  We discuss this in more detail below, including Fig.~\ref{fig:decay}, which shows the covariance between the size of the overshoot and the decay timescale using model $H_{2+p}$.

\section{Slow-down preceding the glitch}
We investigate the frequency slow-down preceding the glitch by extending the $H_{2}$ model to include a step-function in frequency before the glitch. This phenomenologically models a spin-down event sometime before the glitch. The frequency evolution for $H_{2+p}$, the \emph{precursor slow-down model}, is given by
\begin{align}
    f_p(t) = f_2(t) - \Delta f_p \Pi(t; t_g {-} \Delta t, t_g)\,,
    \label{eqn-Hp}
\end{align}
where $f_2(t)$ is given by Eq.~\eqref{eqn-H2}, $\Pi$ is a rectangle function such that the frequency decreases by $\Delta f_p$ for the period $\Delta t$ prior to the glitch. Constructing the model in this way, $\Delta f$ remains the long-term frequency change at the glitch.

Model $H_{2+p}$ is phenomenological and motivated by the data; Fig.~\ref{fig:frequency_evolution_only}. Furthermore, it is one of many simple phenomenological models that could be used; e.g., exponential or linear drift models. Rather than perform a systematic study, we focus solely on $H_{2+p}$ with the aim to motivate further research in this area. To this end, we speculate below about the causes of the slow-down.
Until we have a more physically-grounded model, the significance of the slow-down is difficult to establish.

The Bayes factor shows the precursor slow-down model $H_{2+p}$ is the preferred of all models tested here: comparing with the overshoot-decay $H_{2}$ model (the next most preferred), $\log_{10} B = \bayesf{antiglitch}{two_exponential}$. This suggests the data supports a slow down of the rotation prior to the glitch, in addition to an overshoot and decay.  In Fig.~\ref{fig:frequency_evolution}, we show the best-fit slow-down model in red; the dashed curve represents the raw frequency evolution, and the solid curve the time-averaged best-fit model.

\begin{figure}
    \centering
    \includegraphics[width=\columnwidth]{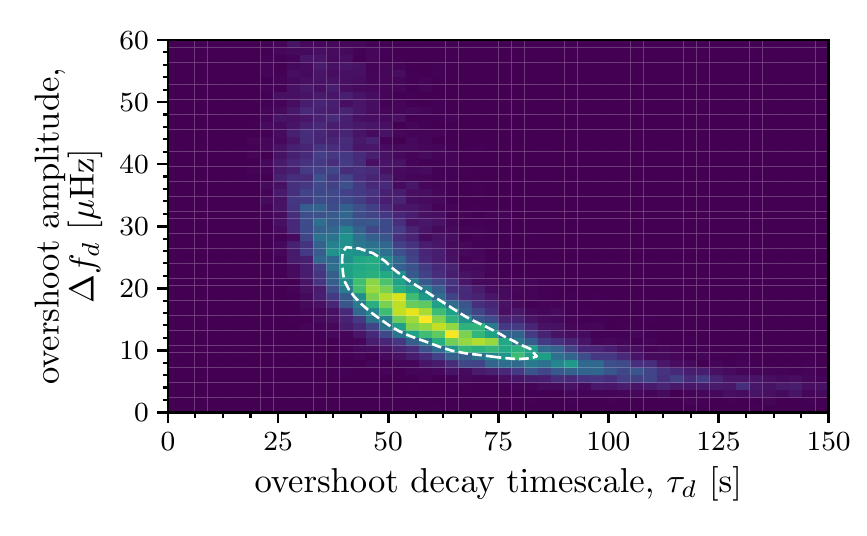}
    \caption{Posterior for overshoot-decay parameters in the $H_{2+p}$ model, corresponding to the red curves in Fig.~\ref{fig:frequency_evolution}.  We show the overshoot decay timescale $\tau_{d}$ and frequency amplitude of the overshoot $\Delta f_{d}$. The white dashed contour indicates the one-sigma confidence level. \label{fig:decay}}
\end{figure}

In Fig.~\ref{fig:decay}, we show the posterior for the size of the overshoot $\Delta f_{d}$ and the overshoot relaxation timescale $\tau_{d}$ using the $H_{2+p}$ model. As mentioned, these are inversely correlated, implying a wide range of equally-likely values for both parameters.

\begin{figure}
    \centering
    \includegraphics[width=\columnwidth]{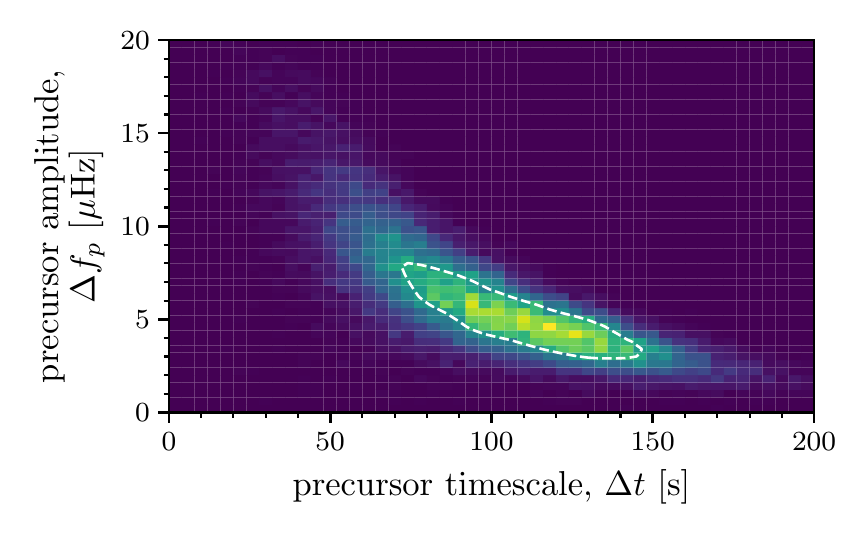}
    \caption{
    Posterior for precursor parameters in the $H_{2+p}$ model, corresponding to the red curves in Fig.~\ref{fig:frequency_evolution}.  We show the precursor time $\Delta t$ and amplitude of the frequency slow-down $\Delta f_p$. The white dashed contour indicates the one-sigma confidence level.  \label{fig:precursor}}
\end{figure}

In Fig.~\ref{fig:precursor}, we show the posterior of the precursor slow-down $\Delta f_p$ and the time before the glitch at which this occurs $\Delta t$. Although the size of the slow-down is not well constrained with $\Delta f_p=\hpdeltafpre$, we note this is a significant fraction of the actual glitch size $\Delta f=\hpdeltaf$ for the $H_{2+p}$ model.

We use a half-normal prior distribution on $\Delta f_p$ (see Table~\ref{tab:priors}, mentary Material), which places the maximum prior probability at zero, gives reasonable support over values $\lesssim \unit[10^{-5}]{Hz}$, and exponentially disfavours larger positive values. If, on the other hand, we use uniform priors, then another local maxima in the posterior distribution becomes present at significantly larger $\Delta f_p$ and at a shorter time preceding the glitch.  We find it difficult to physically motivate a precursor slow-down many times larger than the glitch size; this explains our choice of the half-normal prior.  However, as all our models are phenomenological, we leave open the possibility that this larger mode exists, and leave that for future exploration.

The Vela pulsar has been seen glitching three times with high-time resolution observations\cite{dodson2002,dodson07,palfreyman18}. The first two observations were with a less-sensitive instrument, implying pulses must be folded to obtain sufficient signal-to-noise.  The method detailed herein can be used on those data, and we encourage reanalysis of that data in search of precursor slowdowns and overshoot-decays; such features have been hinted at in Ref.\cite{dodson2002}.

We divide models of the physical nature of the precursor slowdown into two groups.  In the first, the fluctuation seen prior to the glitch in Fig.~\ref{fig:frequency_evolution_only} is a large incarnation of the stochastic fluctuations seen preceding and following the glitch. In the second group, the slow-down is unrelated to this stochastic noise. 

In future work, we will probe the hypothesis that the stochastic fluctuations cause the glitch by developing a statistical analysis of the Vela noise using data away from the glitch. This requires development of a spin-down model in the absence of a glitch that takes into account the stochastic fluctuations, with subsequent analysis of a large number of off-glitch data segments. Comparing the amplitude of the frequency slow-down ($\sim\unit[5]{\mu Hz}$) with the noise distribution, could indicate if the slow-down is a statistical outlier from the typical noise, potentially allowing falsification of the idea that the spin-down event caused the glitch.

The cause of the stochastic fluctuations in Fig.~\ref{fig:frequency_evolution_only} are also of interest. While they could be a manifestation of jitter noise, they may be due to fluctuations in the rotation rate caused by instabilities, or to extrinsic effects such as fluctuations in dispersion measure or scattering\cite{cordes2010}. These ideas can be explored by looking at other sets of high time-resolution data, using the same method used to produce Fig.~\ref{fig:frequency_evolution_only}.

We hypothesise that the slow-down may be due to some intrinsic mechanism in the star, although this is far from certain.  Reference\cite{palfreyman18} reported short-timescale variations in pulse shape during the glitch, including a null pulse and unusual pulse shapes before and after the null.  These could indicate the glitch or preceding slow-down are magnetospheric in origin, although it is difficult to disentangle cause and effect.  Moreover, the large change in spin period on such short timescales is difficult to quantitatively explain without catastrophically changing the magnetic-field topology, which would likely be accompanied with a long-term change in pulse shape as observed following glitches in high magnetic-field pulsars and magnetars\cite{dib2008, weltevrede2011, archibald2016}.  Such a long-term change in pulse shape has not been observed following the 2016 Vela glitch\cite{palfreyman18}.


\section{Did the spin-down event cause the glitch?}
That the glitch is preceded by a spin-down event is intriguing.  Many  models\cite{alpar1981, link1991, glampedakis2009, pizzochero2011, haskell12, graber18} posit that glitches are triggered when sufficient lag is built up between the superfluid component of the inner crust and the lattice crust.  Some fraction of the fluctuations in Fig.~\ref{fig:frequency_evolution_only} may be due to intrinsic, stochastic variations of the star's rotational period. Those variations may take place on timescales faster than the coupling timescales of the crust and internal components. If the slow-down event is one such stochastic variation, albeit a large one, we hypothesise this may trigger the glitch by spinning down the crust and driving the lag above its critical value.  

Our hypothesis has natural corollaries for glitch statistics of the pulsar population: if the stochastic variations are large or comparable to the change in spin period from dipole radiation, glitches will occur probabilistically when the combination of the spin down and variations takes the crust-core lag above the trigger threshold.  In such cases, the time period between glitches would neither be regular nor Poisson-distributed, but would depend on the relative size of the variations with respect to the spin down.  When the variations are large with respect to the spin down, the glitch recurrence time should be Poisson-distributed. Finally, if the variations are small compared to changes from magnetic spin down, the glitch recurrence time should only depend on the spin-down timescale of the system. Additionally, we expect distinct behavior for those stars where glitches are driving the system far from the critical point and spin down is required to return to the threshold before variations can initiate a subsequent glitch, versus the objects where this is not the case and variations can always trigger a glitch.

It is worth noting that, in reality, the critical lag will depend on the neutron-star density and not have a single value, but there is likely also a statistical distribution associated with the macroscopically-averaged value of the critical lag.  This should not significantly effect the arguments presented above as they are mainly qualitative. However it should be taken into account when quantitatively defending this model. 

Although there are two populations of pulsars according to their glitch recurrence statistics\cite{melatos2008, fuentes2017, howitt2018}, more work is required to establish whether the two populations correlate with the relative magnitude of the pulsar's stochastic variations and their spin-down timescales.  Such a statistically rigorous study would potentially be difficult; it is not clear which fluctuations are related to intrinsic pulsar spin noise and, e.g., pulse-jitter noise.  This would also be difficult to generalize to other pulsars.


\section{Conclusion}
During the 2016 glitch, the Vela pulsar first spun down.  A few seconds later it rapidly spun up, before finally spinning down with an exponential relaxation time of $\sim\unit[60]{s}$. This model is substantially favoured over a simple step glitch, or one with only a single spin-up event (see Tab.~\ref{tab:models}). 

Testing the rise time alone, we constrain $\tau_r\le \ninetyrisetime$ (90\% confidence), consistent with the estimated value of Ref.\cite{palfreyman18}, and reducing the previous-best constraint of $\tau_r\lesssim$\unit[30]{s} for the 2004 Vela glitch\cite{dodson07}. Invoking a two-component neutron-star model, our new constraint translates into a lower limit for the mutual-friction coupling of $\B\gtrsim5.7\times10^{-6}$; Eq.~\eqref{eqn:B}.

We find a frequency overshoot and exponential relaxation with amplitude $\Delta f_{d}=\hpdeltafdec$ and decay time scale $\tau_{d}=\hptaudec$ for the $H_{2+p}$ model, a feature that can be theoretically explained\cite{vaneysden2010,haskell12,antonelli17,graber18}. For example, within the three-component model of Ref.\cite{graber18}, the overshoot only exists if the crust mutual-friction coefficient (coupling the crustal superfluid and crust) exceeds the core friction coefficient (coupling the core superfluid and crust). Providing a more qualitative analysis of the internal physics, e.g., constraining moments of inertia and coupling coefficients, is difficult at this point as the phenomenological models studied herein do not produce sufficient information.

Finally, we find evidence for a slow-down, or possibly a precursor \textit{antiglitch}, immediately before the glitch.  To the best of our knowledge this has not been predicted. We hypothesize that it may be a statistical fluctuation consistent with the overall noise fluctuations and speculate such fluctuations drive the differential lag between the superfluid and the crust above its critical value, thus triggering the glitch. This suggests a large number of glitches could be preceded by a slow-down, providing testable predictions.

Analyses like that presented herein only assess the relative evidence of models.  We focus on phenomenological, albeit physically-motivated models, in a bid to remain model agnostic. Even the best fitting models tested here do not explain all the features in the data, e.g., Fig.~\ref{fig:frequency_evolution}.
Future explorations may uncover new descriptions that explain the data better than the models used herein. For example, further theoretical modelling may provide a more nuanced view of how the slow-down preceding the glitch should manifest; the method we developed is easily extendable to compare more complex models.

While direct modelling is one avenue of further investigation, model-agnostic approaches may also yield considerable insight. Figure~\ref{fig:frequency_evolution} is a first step in this direction, although it has the subtlety that the time window distorts temporal and amplitude features.  Another method could be e.g., shapelet-based models for the frequency evolution, providing a means to study the underlying frequency evolution without modelling constraints.

\bibliography{bibliography}

\textbf{Correspondence and requests for materials} should we addressed to G.A, \href{gregory.ashton@ligo.org}{gregory.ashton@ligo.org}.

\textbf{Acknowledgements}
We are grateful to Andrew Melatos, Ian Jones, and the anonymous reviewers for valuable comments. 
Computations were performed on the OzStar supercomputer.
PDL is supported through Australian Research Council Future Fellowship FT160100112 and Discovery Project DP180103155. VG is supported by a McGill Space Institute postdoctoral fellowship and the Trottier Chair in Astrophysics and Cosmology.

\textbf{Contributions to the paper}
G.A. is responsible for the data analysis; G.A., P.D.L., and V.G. are responsible for the model development and discussion; J.P. is responsible for the data collection and reduction.

\textbf{Competing Interests}
Authors declare no competing interests.

\begin{methods}
We use the 72-min stretch of data collected by the University of Tasmania Mt Pleasant 26-m radio telescope\cite{palfreyman18} on 2016 December 12.
The raw flux is analysed fitting a standard pulse template to individual pulses and estimating the site arrival time of each pulse. 
We use {\tt Tempo2}\cite{hobbs2006,edwards2006} to convert site arrival times to solar-system barycentre times, and use the Bayesian analysis package {\tt Bilby}\cite{bilby} to fit timing models.  

The likelihood for the $i$th pulse with observed arrival time $t_i$ is calculated from
\begin{equation}
    \mathcal{L}(t_i; \theta) = \frac{1}{\sqrt{2\pi\sigma^2}} 
                               {\rm exp}\left[-\frac{(t_i - h(i; \theta))^2}{2\sigma^2}\right],
\end{equation}
where $h(i; \theta)$ is the predicted arrival time of the $i$th pulse (within the context of the model). The variance of this distribution is further given by $\sigma^2 = \sigma_i^2 + \sigma_0^2$, where $\sigma_i^2$ is the estimated variance of the $i$th arrival time (as output by the matched-filter profile analysis) and $\sigma_0$ is an additional stochastic noise to be fit for. The priors used are listed in Table~\ref{tab:priors}.

\begin{table}[h]
    \centering
    \begin{tabular}{l|l|l|l}
         &  Prior distribution & Units & Models\\ \hline
         $f_0$ & $\textrm{Uniform}(11.1854, 11.1874)$  & Hz & all\\
         $\phi_0$ & $\textrm{Uniform}(-5, 5)$  & --- & all \\
         $\sigma_0$ & $\textrm{Uniform}(0, 0.01)$ & s & all \\
         $\Delta f$ & $\textrm{Log-Uniform}(10^{-8}, 10^{-4})$ & Hz & all \\
         $t_g$ & $\textrm{Uniform}(-100, 100)$ & s & all \\
         $\tau_r$ & $\textrm{Uniform}(0, 1000)$  & s  & $H_1, H_2, H_{2+p}$\\
         $\Delta f_d$ & $\textrm{Log-Uniform}(10^{-8}, 10^{-4})$ & Hz  & $H_2, H_{2+p}$\\
         $\tau_d$ & $\textrm{Uniform}(0, 1000)$  & s  & $H_2, H_{2+p}$\\
         $\Delta t$ & $\textrm{Uniform}(0, 500)$  & s  & $ H_{2+p}$\\
         $\Delta f_p$ & $\textrm{Half-Normal}(0, 10^{-5})$ & Hz  & $H_{2+p}$\\
    \end{tabular}
    \caption{Table of priors used throughout this work. For the parameters not introduced in the text, $\phi_0$ is the phase parameter (number of rotations), we provide a wider prior to allow the reference pulse to not be zero; and $\sigma$ is the standard-deviation of the Gaussian likelihood.}
    \label{tab:priors}
\end{table}
\end{methods}

\textbf{Data availability}
The data used in this work is available from Ref.\cite{palfreyman18}.

\textbf{Code availability}
The \texttt{bilby}\cite{bilby} analysis code is available from \url{https://git.ligo.org/lscsoft/bilby} and particular scripts for this analysis are available on request from the authors.


\end{document}